\documentclass[12pt]{article}
\usepackage{feynmf}
\usepackage{psfig}

\newcommand{\be}{\begin{equation}}
\newcommand{\ee}{\end{equation}}
\newcommand{\ba}{\begin{eqnarray}}
\newcommand{\ea}{\end{eqnarray}}

\newcommand{\bo}[1]{\mbox{\boldmath $#1$}}

\title{A diagrammatic approach to study the information transfer in 
weakly non-linear channels}

\author{{\bf\sc E. Korutcheva}\\Departamento de F\'{\i}sica Fundamental\\
Universidad Nacional de Educaci\'on a Distancia\\
c/Senda del Rey No 9  - 28080 Madrid, Spain
{\thanks{
electronic address: elka@fisfun.uned.es;\,\
Permanent address: G. Nadjakov Institute of 
Solid State Physics, Bulgarian Academy of Sciences, 1784 Sofia, Bulgaria}}\\
\\{\bf\sc Valeria Del Prete} \\Institute for theoretical physics\\
University of Leuven\\ Celestijnenlaan 200 D, B-3001 Leuven, Belgium}

\date{}
\begin{document}

\bibliographystyle{unsrt}

\maketitle



\newpage

\begin{abstract}
In a recent work we have introduced a novel approach to study the effect of 
weak 
non-linearities in the transfer function on the information transmitted by an 
analogue channel, by means of a perturbative 
diagrammatic expansion. We extend here the analysis to all orders 
in perturbation theory, which allows us to release any constraint concerning 
the 
magnitude of the expansion parameter and to establish the rules to calculate 
easily 
the contribution at any order.
As an example we explicitly compute the information up to the second order in the 
non-linearity, in presence of random gaussian connectivities  
and in the limit when the output noise is not small. We analyze the first and 
second order 
contributions to the mutual information as a function of the non-linearity and of 
the number of output units.
We believe that an extensive application of our method via the analysis of 
the different 
contributions at distinct orders might be able to fill a gap between well known analytical 
results obtained for linear channels and the non trivial treatments which are required 
to study highly non-linear channels.
\end{abstract}

\newpage

\section{Introduction}

Several recent investigations have explored the efficiency of two-layer networks 
in transmitting
information, given that the distribution of the input layer and the input-output 
transformation 
are known \cite{Tre+98b,Tre+95,vale+01c,Guid,vanHerten,NP1,KNP,TKP,vale+01a}.
Some of these works have been inspired by processes involving populations of real 
neurons in the brain \cite{Tre+98b,Tre+95,vale+01c}. There are experimental evidences 
that real neurons operate a highly non-linear transformation of the inputs, whose 
main features can be well captured by a threshold-linear function. Moreover this type of 
transformation allows an easier analytical treatment, at least under the 
assumption of replica symmetry \cite{Tre+98b,Tre+95,vale+01c}.  
A sigmoidal transfer function has been proposed in other works and in 
the more general context of neural networks \cite{Ami+89,Hertz}. 
Yet this choice makes the analytical treatment much more difficult.

In a more theoretical framework, recent studies have explored the communication properties 
of linear and binary channels, where the transfer function is 
either purely linear \cite{Guid,vanHerten}, or highly non-linear 
\cite{NP1,KNP,TKP}.
The analytical solution in case of pure linearity is straightforward, while 
in presence of a step transfer function one must resort to some approximations like 
replica symmetry, or restrict oneself to some particular regions in the parameter space.

No extensive study has been performed yet, trying to bridge these two limit cases, 
pure linearity or strong non-linearity in the transfer function, with respect to 
the impact on the information content, i.e. on the channel efficiency.

In a very recent study \cite{vale+01a}, 
the contribution of small non-linearities to the 
mutual information has been evaluated in case of a gaussian noisy channel, introducing a 
novel approach by means of a perturbative expansion and providing an elegant interpretation 
in terms of feynman diagrams \cite{AGD}.
An analytical expression for the mutual information has been obtained 
at first order in the non-linearity.

Here we extend the previous analyses to all orders in perturbation theory, deriving both 
the analytical expansion and the diagrammatic formalism necessary to evaluate the 
contribution to the mutual information at each order.
Then we apply our method and 
quantify the first and second order contributions to the mutual information in the 
case of random gaussian connectivities and in the 
limit of large output noise.
  
Even though not motivated by any particular hardware or biological application, 
our study is an attempt to fill a theoretical gap between 
purely linear and strongly non-linear channels, 
at least in the case of gaussian units.

We believe that an extensive application of our expansion will allow to examine 
the impact of 
non-linearities on the information transmission and its modulation with the noise and 
the other parameters in the model.
This will be the object of future investigations.

\section{The model}
The network model is analogous to the one used in \cite{vale+01a}.

The distribution of the $N$ continuous inputs
$\bo x$=$\{x_1...x_N\}$ is gaussian with correlation 
matrix $C$, and each input signal is corrupted by uncorrelated 
gaussian noise $\bo \nu$=\{$\nu_1..\nu_N\}$,
as in the following:

\begin{eqnarray}
\langle x_i\rangle&=&0,\\
\langle x_ix_j\rangle&=&[C]_{ij}, \,\,\,\, \forall i,j\in 1,2,..N.
\label{input}
\end{eqnarray} 

\begin{eqnarray}
\langle \nu_i\rangle&=&0,\\
\langle \nu_i\nu_j\rangle&=&b_0\,\delta_{ij}, \,\,\,\, \forall i,j\in 1,2,..N.
\label{inputnoise}
\end{eqnarray}

Each output unit performs a linear summation of the noisy inputs $\bo x+\bo \nu$ 
via the matrix of the connectivities $J=\{J_{ij}\}$; the result is transformed via a non-linear 
transfer function $G$:

\begin{equation} 
\bo V=G(J(\bo x+\bo \nu))+\bo z,
\label{output}
\end{equation}
where $\bo z$=\{$z_1..z_N\}$ is uncorrelated gaussian noise affecting the output,
according to the following distribution:

\begin{eqnarray}
\langle z_i\rangle&=&0,\\ 
\langle z_iz_j\rangle&=&b\,\delta_{ij}, \forall i,j\in 1,2,..N.
\label{outputnoise}
\end{eqnarray}

In analogy to the case examined in \cite{vale+01a}, we consider a sigmoidal 
shaped transfer function assuming that the non-linearity is small, so that 
a Taylor expansion can be performed:

\begin{equation}
G(x)=th( x)=\sum_{k=0}^\infty (-1)^k \frac{x^{2k+1}}{2k+1}.
\label{transfer}
\end{equation}
For small values of $x$ one can stop at the first non-linear term, the cubic one.
To illustrate the method we will first focus on a cubic non-linearity for the sake 
of simplicity. Nonetheless, 
as it will be clear in the following, out method is equally applicable to any term 
in the expansion of the hyperbolic tangent, finally allowing to reconstruct the whole 
series.

In a cubic approximation eq.(\ref{output}) can be rewritten as:
 
\begin{equation} 
\bo V\simeq \bo h + \bo g(\bo h)+ \bo z,
\label{output2}
\end{equation}
with:
\begin{eqnarray} 
\bo g(\bo h)=g_0 \bo h^3,\nonumber\\
\bo h=J(\bo x+\bo\nu).
\end{eqnarray}

\section{A perturbative approach to the mutual information}
In analogy to what is reported in \cite{vale+01a} we express the mutual information 
between input and output as the difference between the {\it output entropy} and the 
{\it equivocation } between input and output\footnote{Using the natural logarithm we 
implicitly measure the information in natural numbers. Conversion to bits is easily 
obtained dividing the mutual information by $\ln 2$:}:

\begin{equation}
I=\int d\bo x\int d\bo V P(\bo x,\bo V)\ln \frac{P(\bo x,\bo V)}{P(\bo x)P(\bo V)}=
H(\bo V)-\langle H(\bo V|\bo x)\rangle_{x};
\label{info}
\end{equation}

\begin{equation}
H(\bo V)=-\int d\bo V P(\bo V)\ln P(\bo V)
\label{entropy}
\end{equation}

\begin{equation}
\langle H(\bo V|\bo x)\rangle_{x}= -\int d\bo x \int d\bo V P(\bo x)P(\bo V|\bo x)\ln P(\bo V|\bo x).
\label{equi}
\end{equation}

The probabilities $P(\bo V)$, $P(\bo x)$, $P(\bo V|\bo x)$ are fixed by 
eqs.(\ref{input}),
(\ref{inputnoise}),(\ref{output2}); in particular explicit expressions for 
both $P(\bo V)$ and $P(\bo V|\bo x)$ can be obtained explicitating the dependence of 
$\bo V$ on $\bo h$. One has:

\begin{eqnarray}
P(\bo V)&=&\int d\bo h P(\bo h) P(\bo V|\bo h);\\
P(\bo V|\bo x)&=&\int d\bo h P(\bo h|\bo x) P(\bo V|\bo h);\\
P(\bo V|\bo h)&=&\frac{1}{\sqrt{2\pi b^N}}\cdot 
e^{-{({\bf V}-{\bf h}-{\bf g}({\bf h}))}^2/2b}.
\label{probs}
\end{eqnarray}
where $\bo h$=$J(\bo x+\bo \nu)$. Given the relationships (\ref{input}),
(\ref{inputnoise}) it is trivial to derive that both $P(\bo h)$ and 
$P(\bo h|\bo x)=P(\bo h-J\bo x)$ are gaussian distributed:

\be
P(\bo h)=\frac{e^{-\frac{1}{2}\bf h\left(A\right)^{-1}\bf h}}
{\sqrt{(2\pi)^NdetA}};\,\,\,A=JCJ^T+b_0JJ^T,
\label{p_h}
\ee

\be
P(\bo h|\bo x)=\frac{e^{-\frac{1}{2}\left(\bf h-J\bf x\right)\left(B\right)^{-1}
\left(\bf h-J\bf x\right)}}
{\sqrt{(2\pi)^NdetB}};\,\,\,B=b_0JJ^T.
\label{p_h_x}
\ee
It is just the presence of only gaussian averages that will finally allow us to 
express the mutual information as a series of Feynman diagrams by means 
of the Wick theorem.

Further details can be found in \cite{vale+01a}.

\subsection{Perturbative expansion for the output entropy}

Let us consider eq.(\ref{entropy}). When the non-linear term $\bo g(\bo h)$ is non zero, 
integration on $\bo h$ cannot be performed without resorting to some approximation.
In \cite{vale+01a} it has been shown that an expansion up to first order in 
$\bo g(\bo h)$ allows to perform the integration and derive an analytical 
expression for the mutual information. 
Our purpose here is to extend the previous analysis, considering all terms in the 
expansion, of whatever order in $\bo g(\bo h)$.

Let us consider the following equalities:

\begin{equation}
P(\bo V)=\int d\bo h P(\bo h) P_0(\bo V|\bo h) e^{\Delta/b}=
\int d\bo h P(\bo h) P_0(\bo V|\bo h) 
\left[\sum_{l=0}^\infty \frac{\Delta^l}{l!\,b^l}\right],
\label{p_v_exp}
\end{equation}
where
\begin{eqnarray}
P_0(\bo V|\bo h)&=&\frac{1}{\sqrt{2\pi b^N}}\cdot 
e^{-{({\bf V}-{\bf h})}^2/2b},\label{p_0_v_h}\\ 
\Delta &=&\Delta\left(\bo V,\bo h\right)=
\left(\bo V-\bo h\right)\cdot\bo g(\bo h)-\frac{\bo g(\bo h)^2}{2}
\label{delta}
\end{eqnarray}
and we have performed the perturbation expansion in terms of $\Delta$, 
keeping in mind that an explicit expression in terms of powers of $\bo g$ 
can be extracted a posteriori.

Inserting the expansion in powers of $\Delta$, eq.(\ref{p_v_exp}), in eq.(\ref{entropy}) 
it can be shown that the output entropy can be expressed as follows:
\begin{eqnarray}
H(\bo V)&=&H_0(\bo V)-\int d\bo V\left(\sum_{l=1}^\infty 
\frac{\left\langle \Delta^l\right\rangle_h}{l!\,b^l}\right)
\ln P_0\left(\bo V\right)\nonumber\\
&+&\int d\bo V P_0(\bo V)\sum_{n=2}^{\infty} \frac{(-1)^{n+1}}{n(n-1)}
\left(\sum_{l=1}^\infty \frac{1}{P_0(\bo V)}
\frac{\left\langle \Delta^l\right\rangle_h}{l!\, b^l}\right)^n,
\label{final_entropy}
\end{eqnarray}
where we have used the notation:
\begin{equation}
\left\langle \Delta^l\right\rangle_h=
\int d\bo h P(\bo h)P_0(\bo V|\bo h)\Delta^l
=\int d\bo h P(\bo h)P_0(\bo V|\bo h)
\left[\left(\bo V-\bo h\right)\cdot\bo g(\bo h)-\frac{\bo g(\bo h)^2}{2}\right]^l.
\label{delta2}
\end{equation}
$H_0(\bo V)$ is the output entropy when $\bo g=0$:
\be
H_0(\bo V)=\int d\bo V P_0(\bo V)\ln P_0(\bo V).
\ee
From eqs.(\ref{p_0_v_h}),(\ref{p_h}) it is easy to derive that:
\be
P_0(\bo V)=\int d\bo h P(\bo h) P_0(\bo V|\bo h)
\frac{e^{-\frac{1}{2}\bf V\left(A+bI\right)^{-1}\bf V}}{\sqrt{(2\pi)^Ndet(A+bI)}},
\label{p_0_v}
\ee
which allows to derive an explicit expression for $H_0(\bo V)$:
\be
H_0(\bo V)=\frac{N}{2}\left[1+\ln 2\pi\right]+\frac{1}{2}det(A+bI).
\label{H_0_v}
\ee

In the limit when $g_0$ is very small and $g_0\ll b$ one can stop at the first order in
$\Delta$, neglecting the second order term in $\bo g$. 
The expression for the output entropy becomes:

\begin{equation}
H(\bo V)\simeq H_0(\bo V) - \frac{1}{b}\int d\bo h\int d\bo V P(\bo h) 
P_0(\bo V|\bo h) \bo g(\bo h)(\bo V- \bo h) \ln P_0(\bo V).
\label{outfin}
\end{equation}
 
In this simpler case, as it has been shown in \cite{vale+01a}, integration is 
straightforward, leading to an explicit final expression for the output entropy, 
as well as for the mutual 
information.
Here we will show that it is possible to generalize our approach to every order 
in perturbation theory, establishing the basic rules to identify each integral
with a diagram, so that the analytical evaluation is reduced to a combinatorial problem 
via the use of Wick theorem \cite{AGD}.

Let us reconsider eq.(\ref{delta2}). A change of variables allows to reduce the 
integration on $\bo h$ to a gaussian:
\ba 
\bo h&\longrightarrow &\bo\gamma+\frac{1}{b}D\bo V;\nonumber\\
\left\langle \Delta^l\right\rangle_h\!\!\!&\!\!\!\longrightarrow&P_0(\bo V)\int d\bo\gamma 
P(\bo \gamma)\left[g_0\left(\bo V-\bo\gamma-\frac{1}{b}D\bo V\right)
\left(\bo\gamma+\frac{1}{b}D\bo V\right)^3
-\frac{g_0^2}{2}\left(\bo\gamma+\frac{1}{b}D\bo V\right)^6\right]^l
\ea
where
\be
P(\bo\gamma)=
\frac{e^{-\frac{1}{2}\bo\gamma D^{-1}\bo\gamma}}{\sqrt{(2\pi)^{N}detD}};\,\,\,D^{-1}=A^{-1}+b^{-1}I.
\label{p_gamma}
\ee

Inserting this expression in eq.(\ref{final_entropy}) one obtains:

\begin{eqnarray}
&&H(\bo V)=H_0(\bo V)+\nonumber\\
&&\frac{1}{2}\left\langle\left\langle
\sum_{l=1}^\infty \frac{g_0^l}{l!\,b^l}
\left[\left(\bo V-\bo\gamma-\frac{1}{b}D\bo V\right)
\left(\bo\gamma+\frac{1}{b}D\bo V\right)^3
-\frac{g_0}{2}\left(\bo\gamma+\frac{1}{b}D\bo V\right)^6\right]^l\right\rangle_\gamma
\left(\bo V\left[A+bI\right]^{-1}\bo V\right)\right\rangle_V\nonumber\\
&&+\left\langle\sum_{n=2}^{\infty} \frac{(-1)^{n+1}}{n(n-1)}
\left\langle\sum_{l=1}^\infty \frac{g_0^l}{l\!\,b^l}\left[\left(\bo V-\bo\gamma-\frac{1}{b}D\bo V\right)
\left(\bo\gamma+\frac{1}{b}D\bo V\right)^3
-\frac{g_0}{2}\left(\bo\gamma+\frac{1}{b}D\bo V\right)^6\right]^l\right\rangle_\gamma^n
\right\rangle_V,
\label{final_entropy2}
\end{eqnarray}
where 
\begin{eqnarray}
\langle F(\bo \gamma,\bo V)\rangle_\gamma&=
&\int d\bo \gamma P(\bo \gamma)F(\bo \gamma,\bo V),\nonumber\\
\langle F(\bo \gamma,\bo V)\rangle_V&=
&\int d\bo V P_0(\bo V)F(\bo \gamma,\bo V).
\end{eqnarray}
$P(\bo\gamma)$,$P_0(\bo V)$ are given respectively by eqs.(\ref{p_gamma}), 
(\ref{p_0_v}) and we have explicitely put $\bo g(\bo h)=g_0\bo h^3$.

\subsection{Perturbative expansion for the equivocation}
A very similar expression can be obtained for the equivocation, 
by means of the same perturbative approach.

Let us reconsider eqs.(\ref{equi}),(\ref{probs}).
Since $P(\bo V|\bo x)$ can be expressed as in integral on the distribution
$P(\bo V|\bo h)$, we can use the same perturbative expansion already introduced
for $P(\bo V)$ in eq.(\ref{p_v_exp}):

\begin{equation}
P(\bo V|\bo x)=\int d\bo h P(\bo h|\bo x) P_0(\bo V|\bo h) e^{\Delta/b}=
\int d\bo h P(\bo h|\bo x) P_0(\bo V|\bo h) 
\left[\sum_{l=0}^\infty \frac{\Delta^l}{l !\,b^l}\right],
\label{p_v_x_exp}
\end{equation}
where $P_0(\bo V|\bo h)$ and $\Delta$ are given in 
eqs.(\ref{p_0_v_h}),(\ref{delta}) and 
$P(\bo h|\bo x)$ is given in eq.(\ref{p_h_x}).

Introducing eq.(\ref{p_v_x_exp}) in the expression for the equivocation one obtains:
\begin{eqnarray}
\left\langle H(\bo V|\bo x)\right\rangle_x&=&\left\langle H_0(\bo V|\bo x)\right\rangle_x
-\int d\bo x P(\bo x)\int d\bo V\left(\sum_{l=1}^\infty 
\frac{\left\langle \Delta^l\right\rangle_{h|x}}{l !\,b^l}\right)\ln P_0\left(\bo V|\bo x)
\right)\nonumber\\
&+&\int d\bo x\int d\bo V P_0(\bo V|\bo x)\sum_{n=2}^{\infty} \frac{(-1)^{n+1}}{n(n-1)}
\left(\sum_{l=1}^\infty \frac{1}{P_0(\bo V|\bo x)}
\frac{\left\langle \Delta^l\right\rangle_{h|x}}{l! \,b^l}\right)^n,
\label{final_equi}
\end{eqnarray}
where we have used the notation:
\be
\left\langle \Delta^l\right\rangle_{h|x}=
\int d\bo h P(\bo h|\bo x)P_0(\bo V|\bo h)\Delta^l
=\int d\bo h P(\bo h|\bo x)P_0(\bo V|\bo h)
\left[\left(\bo V-\bo h\right)\cdot\bo g(\bo h)-\frac{\bo g(\bo h)^2}{2}\right]^l
\label{delta2_x}
\ee
and $\left\langle H_0(\bo V|\bo x)\right\rangle_x$ is the equivocation when $\bo g=0$:

\be
\left\langle H_0(\bo V|\bo x)\right\rangle_x=
\int d\bo x\int d\bo V P(\bo x)P_0(\bo V|\bo x)\ln P_0(\bo V|\bo x),
\ee

\be
P_0(\bo V|\bo x)=\frac{e^{-\frac{1}{2}\left(\bf V-J\bf x\right)\left(B+bI\right)^{-1}
\left(\bf V-J\bf x\right)}}
{\sqrt{(2\pi)^Ndet(B+bI)}}.
\label{p_V_x}
\ee
Then it is easy to derive the final expression for the zeroth order equivocation:

\be
\left\langle H_0(\bo V|\bo x)\right\rangle_x=
\frac{N}{2}\left(1+\ln 2\pi\right)+\frac{1}{2}\ln det\left(B+bI\right).
\label{H_0_v_x}
\ee

On the other hand, if one keeps the contributions to the information up to first 
order in $g_0$ one obtains:

\begin{equation}
\langle H(\bo V|\bo x)\rangle_{x}\!\simeq\! 
\langle H_0(\bo V|\bo x)\rangle_{x} - 
\frac{1}{b}\int\!\! d\bo h\!\!\int \!\!d\bo x\!\!\int\!\! 
d\bo V \!P(\bo x) P(\bo h|\bo x)P_0(\bo V|\bo h) \bo g(\bo h)(\bo V\!-\! \bo h) 
\ln P_0(\bo V|\bo x).
\label{eqfin}
\end{equation}

Eq.(\ref{final_equi}) can be treated in a very similar way as already done in the 
case of the output entropy, eq.(\ref{final_entropy}).
Since three different integrations are present in eq.(\ref{final_equi}) diagonalization 
of the Gaussian distributions requires replacing both 
$\bo h$ and $\bo V$ in sequence:

\ba
\bo h&\longrightarrow&\tilde{\bo \gamma}+
\left(\frac{\bo V}{b}+B^{-1}J\bo x\right)\tilde{D},\,\,\,\,\,
\tilde{D}^{-1}=B^{-1}+b^{-1}I,\nonumber\\
\bo V&\longrightarrow&\bo u+J\bo x.
\ea

By means of these substitutions eq.(\ref{final_equi}) can be rewritten as follows:

\begin{eqnarray}
&&\left\langle H(\bo V|\bo x)\right\rangle_x=
\left\langle H_0(\bo V|\bo x)\right\rangle_x+\nonumber\\
&&\frac{1}{2}\left\langle\left\langle
\sum_{l=1}^\infty \frac{g_0^l}{l!\,b^l}
\left[\left(\bo u\!-\!\tilde{\bo\gamma}\!-\!\frac{1}{b}\tilde{D}\bo u\right)\!\!\!
\left(\tilde{\bo\gamma}\!+\!\frac{1}{b}\tilde{D}\bo u\!+\!J\bo x\right)^3
-\!\!\!\frac{g_0}{2}\left(\tilde{\bo\gamma}\!+\!\frac{1}{b}\tilde{D}\bo u\!
+\!J\bo x\right)^6\right]^l
\right\rangle_{\tilde{\gamma},x}\!\!\!
\left(\bo u\left[B+bI\right]^{-1}\bo u\right)\right\rangle_u\nonumber\\
&&+\left\langle\sum_{n=2}^{\infty}\frac{(-1)^{n+1}}{n(n-1)}
\left\langle\sum_{l=1}^\infty \frac{g_0^l}{l!\,b^l}
\left[\left(\bo u\!-\!\tilde{\bo\gamma}\!-\!\frac{1}{b}\tilde{D}\bo u\right)
\!\left(\tilde{\bo\gamma}\!+\!\frac{1}{b}\tilde{D}\bo u\!+\!J\bo x\right)^3\!
-\!\frac{g_0}{2}\left(\tilde{\bo\gamma}\!+\!\frac{1}{b}\tilde{D}\bo u\!
+\!J\bo x\right)^6\right]^l
\right\rangle_{\tilde{\gamma}}^n
\right\rangle_{u,x}
\label{final_equi2}
\end{eqnarray}
where 
\begin{eqnarray}
\langle F(\tilde{\bo \gamma},\bo u,\bo x)\rangle_{\tilde{\gamma}}&=&\int d\tilde{\bo \gamma} P(\tilde{\bo \gamma})F(\tilde{\bo \gamma},\bo u,\bo x),\nonumber\\
\langle F(\tilde{\bo \gamma},\bo u,\bo x)\rangle_u&=&\int d\bo u P(\bo u)F(\tilde{\bo \gamma},\bo u,\bo x),
\end{eqnarray}
\be
P(\tilde{\bo\gamma})=
\frac{e^{-\frac{1}{2}\tilde{\bo\gamma} \tilde{D}^{-1}\tilde{\bo\gamma}}}
{\sqrt{(2\pi)^{N}det\tilde{D}}},
\label{p_t_gamma}
\ee
\be
P(\bo u)=
\frac{e^{-\frac{1}{2}\bf u (B+bI)^{-1}\bf u}}{\sqrt{(2\pi)^{N}det(B+bI)}}.
\label{p_u}
\ee

As it is evident from eqs.(\ref{final_entropy2}),(\ref{final_equi2}) the mutual information 
is expressed as series of Gaussian averages, where all powers higher than the second one 
can be treated via the Wick theorem.
This allows to establish the basic rules for a diagrammatic expansion in terms of 
Feynman graphs, which is the subject of the next sections.

\section{A diagrammatic formalism for the expansion}

It is well known that higher order moments of a Gaussian distribution can be 
reduced to a series of products of second order moments via the use of Wick's theorem, 
both in classical and in quantum systems \cite{AGD}.
The introduction of a diagrammatic formalism allows to associate a graph to each type 
of integral. Therefore the whole series can be expressed in a very compact and elegant way 
and integrations can be performed symbolically contracting all the 
lines pair by pair, in such a way to obtain all the topologically distinct diagrams.
Since each contraction is accociated to a precise numerical value, 
the value of each diagram can be easily calculated by simply multiplying all the 
factors corresponding to the contractions of the lines.

In the previous work \cite{vale+01a}, 
where we have focused on the first order terms in $\bo g$, 
in eqs.(\ref{final_entropy2}),(\ref{final_equi2}),
we have identified the basic elements 
which might allow to build a diagrammatic expansion for the mutual information, 
up to first order.
Here we generalize our approach and we show that a diagrammatic interpretation can be 
provided for any order in the expansion, establishing the basic elements and rules distinctly 
for the output entropy and for the equivocation. 

\subsection{Diagrammatic rules for the evaluation of the output entropy}

Since two distinct averages characterize eq.(\ref{final_entropy2}), namely on $\bo \gamma$ and $\bo V$, one is naturally tempted to introduce two 
distinct symbolic lines for $\bo \gamma$ and $\bo V$. Yet $\bf two$ distinct objects are to be contracted and integrated on $\bo V$, namely $\bo V$ itself and $D\bo V$.
Therefore it is more convenient to 
introduce two distinct lines for $\bo V$ and $D\bo V$.
Since either of this two lines can be contracted with itself or with the other one, one has three distinct rules on contraction corresponding to integrating different objects on $\bo V$.
The whole prescription can be given as follows:

\begin{fmffile}{elka1}

\begin{enumerate}
\item Each term $V_i$ is represented by a straight line \,\,\,\,\,\ 
\begin{fmfgraph*}(20,1)\fmfpen{thin}\fmfleft{v1}\fmfright{o1}
\fmflabel{$i$}{v1}\fmfdot{v1}\fmf{plain,label=$V_i$}{v1,o1}\end{fmfgraph*}

\item Each term $(D\bo V)_i$ is represented by a crossed solid line  \,\,\,\,\,\,\ 
\begin{fmfgraph*}(20,1)\fmfpen{thin}\fmfleft{v1}\fmfright{o1}
\fmflabel{$i$}{v1}\fmfdot{v1}\fmf{plain}{v1,o2}
\fmfv{decor.shape=cross,decor.size=3mm}
{o2}\fmf{plain,label=$(D\bo V)_i$}{o2,o1}\end{fmfgraph*}

\item Each term $\gamma_i$ is represented by a wiggly line \,\,\,\,\,\ 
\begin{fmfgraph*}(20,1)\fmfpen{thin}\fmfleft{v1}\fmfright{o1}
\fmflabel{$i$}{v1}\fmfdot{v1}\fmf{wiggly,label=$\gamma_i$}{v1,o1}\end{fmfgraph*}

\item Each matrix element $[A+bI]_{i j}^{-1}$ is represented by a dashed romboid \,\,\,\,\,\,\,\,\ \begin{fmfgraph*}(20,5)
\fmfpen{thin}\fmfleft{i1}\fmfright{o1}
\fmfpoly{shade,tension=0.5}{k1,k2,k3,k4}
\fmflabel{$i$}{k1}\fmflabel{$j$}{k3}\fmf{phantom}{i1,k1}
\fmf{phantom}{k3,o1}\end{fmfgraph*}

\item The integration of the product $V_iV_j$ over $\bo V$ 
corresponds to the contraction of two straight 
lines coming out of vertices $i$,$j$.

This produces a matrix element
$(A+bI)_{i j}$ \,\,\,\,\,\,\,\
\begin{fmfgraph*}(20,1)\fmfpen{thin}\fmfleft{v1}\fmfright{v2}\fmfdot{v1}
\fmflabel{$i$}{v1}
\fmfdot{v2}\fmflabel{$j$}{v2}\fmf{plain}{v1,v2}\end{fmfgraph*}.

\item The integration of the product $V_i(D\bo V)_j/b$ over $\bo V$
corresponds to the contraction of a straight line 
coming out of the vertex $i$ with a straight crossed line coming out of vertex 
$j$.

This produces 
a matrix element $[(A+bI)D]_{i j}/b$ \,\,\,\,\,\,\ \begin{fmfgraph*}(20,1)\fmfpen{thin}
\fmfleft{v1}\fmfright{v2}\fmfdot{v1}
\fmflabel{$i$}{v1}\fmf{plain}{v1,o2}\fmfv{decor.shape=cross,decor.size=3mm}{o2}\fmfdot{v2}
\fmflabel{$j$}{v2}\fmf{plain}{o2,v2}
\end{fmfgraph*}.

\item The integration of the product $(D\bo V)_i(D\bo V)_j/b^2$ over $\bo V$
corresponds to the contraction of two straight crossed line 
coming out of the vertices $i$,$j$.

This produces a matrix element $[D(A+bI)D^T]_{i j}/b^2$ \,\,\,\,\,\,\ 
\begin{fmfgraph*}(20,1)\fmfpen{thin}
\fmfleft{v1}\fmfright{v2}\fmfdot{v1}
\fmflabel{$i$}{v1}\fmf{plain}{v1,o2}\fmfv{decor.shape=cross,decor.size=3mm}{o2}
\fmf{plain}{o2,o3}\fmfv{decor.shape=cross,decor.size=3mm}{o3}
\fmfdot{v2}
\fmflabel{$j$}{v2}\fmf{plain}{o3,v2}
\end{fmfgraph*}.

\item The integration of the product $\gamma_i\gamma_j$ 
over $\bo \gamma$ corresponds to the contraction of two wiggly 
lines coming out of vertices $i$,$j$.

This produces a matrix element 
$D_{i j}$ \,\,\,\,\,\,\,\
\begin{fmfgraph*}(20,1)\fmfpen{thin}\fmfleft{v1}\fmfright{v2}\fmfdot{v1}
\fmflabel{$i$}{v1}
\fmfdot{v2}\fmflabel{$j$}{v2}\fmf{photon}{v1,v2}\end{fmfgraph*}.

\end{enumerate}

Eq.(\ref{final_entropy2}) can be written in terms of these symbols:

\begin{eqnarray}
&&H(\bo V)=H_0(\bo V)+\frac{1}{2}\left\langle\left\langle\left\langle
\sum_{l=1}^\infty \frac{g_0^l}{l!\,b^l}\sum_{ijk}
\left[\left(\quad\parbox{10mm}{\begin{fmfgraph*}(10,1)\fmfkeep{V}\fmfpen{thin}
\fmfleft{v1}
\fmfright{o1}\fmflabel{$i$}{v1}\fmfdot{v1}\fmf{plain}{v1,o1}\end{fmfgraph*}}
-\quad\parbox{10mm}{\begin{fmfgraph*}(10,1)\fmfkeep{gamma}\fmfpen{thin}
\fmfleft{v1}\fmfright{o1}
\fmflabel{$i$}{v1}\fmfdot{v1}\fmf{photon}{v1,o1}\end{fmfgraph*}}-\quad\parbox{10mm}{
\begin{fmfgraph*}(10,1)\fmfkeep{dv}\fmfpen{thin}\fmfleft{v1}\fmfright{o1}
\fmflabel{$i$}{v1}\fmfdot{v1}\fmf{plain}{v1,o2}\fmfv{decor.shape=cross,decor.size=3mm}{o2}
\fmf{plain}{o2,o1}\end{fmfgraph*}}\right)\right.\right.\right.\right.\nonumber\\
&&\left.\left.\left.\left.\left(\,\,\,\parbox{10mm}{\fmfreuse{gamma}}
+\quad\parbox{10mm}{\fmfreuse{dv}}\right)^3
-\frac{g_0}{2}\left(\quad\parbox{10mm}{\fmfreuse{gamma}}+\quad\parbox{10mm}{\fmfreuse{dv}}
\right)^6\right]^l\right\rangle_{\parbox{5mm}{\begin{fmfgraph*}(5,1)\fmfkeep{gamma_av}
\fmfpen{thin}\fmfleft{v1}\fmfright{o1}
\fmfdot{v1}\fmf{photon}{v1,o1}\end{fmfgraph*}}} 
\left(\parbox{15mm}{\begin{fmfgraph*}(15,5)
\fmfpen{thin}\fmfleft{i1}\fmfright{o1}
\fmfpoly{shade,tension=0.5}{k1,k2,k3,k4}
\fmf{plain,label=$j$,label.side=right}{i1,k1}
\fmf{plain,label=$k$,label.side=right}{k3,o1}\end{fmfgraph*}}\right)
\right\rangle_{\parbox{5mm}{\begin{fmfgraph*}(5,1)\fmfkeep{V_av}
\fmfpen{thin}\fmfleft{v1}\fmfright{o1}
\fmfdot{v1}\fmf{plain}{v1,o1}\end{fmfgraph*}}}\,\,\right\rangle_{\parbox{5mm}
{\begin{fmfgraph*}(5,1)\fmfkeep{dv_av}
\fmfpen{thin}\fmfleft{v1}\fmfright{o1}
\fmfdot{v1}\fmf{plain}{v1,o2}\fmfv{decor.shape=cross,decor.size=3mm}{o2}\fmf{plain}{o2,o1}
\end{fmfgraph*}}}\nonumber\\
&&+\left\langle\left\langle\sum_{n=2}^{\infty} \frac{(-1)^{n+1}}{n(n-1)}
\left\langle\sum_{l=1}^\infty \frac{g_0^l}{l!\,b^l}\sum_i
\left[\left(\quad\parbox{10mm}{\fmfreuse{V}}-\quad\parbox{10mm}{\fmfreuse{gamma}}
-\quad\parbox{10mm}{\fmfreuse{dv}}\right)\right.\right.\right.\right.\nonumber\\
&&\left.\left.\left.\left.\left(\quad\parbox{10mm}{\fmfreuse{gamma}}+
\quad\parbox{10mm}{\fmfreuse{dv}}\right)^3
-\frac{g_0}{2}\left(\quad\parbox{10mm}{\fmfreuse{gamma}}+\quad
\parbox{10mm}{\fmfreuse{dv}}\right)^6\right]^l
\right\rangle_{\parbox{5mm}{\fmfreuse{gamma_av}}}^n\,\,
\right\rangle_{\parbox{5mm}{\fmfreuse{V_av}}}\,\,
\right\rangle_{\parbox{5mm}{\fmfreuse{dv_av}}},
\label{entropy_diagr}
\end{eqnarray}
where we have symbolically separated the average across 
$\bo V$ from the average across $D\bo V$ to remind that contractions
have to be performed on both objects according to the rules given above.

\subsection{Diagrammatic rules for the evaluation of the equivocation}

A diagrammatic interpretation can be given for eq.(\ref{final_equi2}) introducing proper
symbols and rules for the contractions:

\begin{enumerate}
\item Each term $u_i$ is represented by a double straight line \,\,\,\,\,\ 
\begin{fmfgraph*}(20,1)\fmfpen{thin}\fmfleft{v1}\fmfright{o1}
\fmflabel{$i$}{v1}\fmfdot{v1}\fmf{dbl_plain,label=$u_i$}{v1,o1}\end{fmfgraph*}

\item Each term $(\tilde{D}\bo u)_i/b$ is represented by a crossed double straight line  \,\,\,\,\,\,\ 
\begin{fmfgraph*}(20,1)\fmfpen{thin}\fmfleft{v1}\fmfright{o1}
\fmflabel{$i$}{v1}\fmfdot{v1}\fmf{dbl_plain}{v1,o2}
\fmfv{decor.shape=cross,decor.size=3mm}
{o2}\fmf{dbl_plain,label=$(\tilde{D}\bo u)_i/b$}{o2,o1}\end{fmfgraph*}

\item Each term $\tilde{\gamma}_i$ is represented by a double wiggly line \,\,\,\,\,\ 
\begin{fmfgraph*}(20,1)\fmfpen{thin}\fmfleft{v1}\fmfright{o1}
\fmflabel{$i$}{v1}\fmfdot{v1}\fmf{dbl_wiggly,label=$\tilde{\gamma}_i$}{v1,o1}\end{fmfgraph*}

\item Each term $(J\bo x)_i$ is represented by a dotted line \,\,\,\,\,\ 
\begin{fmfgraph*}(20,1)\fmfpen{thin}\fmfleft{v1}\fmfright{o1}
\fmflabel{$i$}{v1}\fmfdot{v1}\fmf{dots,label=$(J\bo x)_i$}{v1,o1}\end{fmfgraph*}

\item Each matrix element $[B+bI]_{i j}^{-1}$ is represented by a filled romboid \,\,\,\,\,\,\,\,\ \begin{fmfgraph*}(20,5)
\fmfpen{thin}\fmfleft{i1}\fmfright{o1}
\fmfpoly{full,tension=0.5}{k1,k2,k3,k4}
\fmflabel{$i$}{k1}\fmflabel{$j$}{k3}\fmf{phantom}{i1,k1}
\fmf{phantom}{k3,o1}\end{fmfgraph*}

\item The integration of the product $u_iu_j$ over $\bo u$  
corresponds to the contraction of two double straight 
lines coming out of vertices $i$,$j$.

This produces a matrix element
$(B+bI)_{i j}$ \,\,\,\,\,\,\,\
\begin{fmfgraph*}(20,1)\fmfpen{thin}\fmfleft{v1}\fmfright{v2}\fmfdot{v1}
\fmflabel{$i$}{v1}
\fmfdot{v2}\fmflabel{$j$}{v2}\fmf{dbl_plain}{v1,v2}\end{fmfgraph*}.

\item The integration of the product $u_i(\tilde{D}\bo u)_j/b$ over $\bo u$
corresponds to the contraction of a double straight line 
coming out of the vertex $i$ with a double straight crossed line coming out of vertex 
$j$.

This produces 
a matrix element $B_{i j}$ \,\,\,\,\,\,\ \begin{fmfgraph*}(20,1)\fmfpen{thin}
\fmfleft{v1}\fmfright{v2}\fmfdot{v1}
\fmflabel{$i$}{v1}\fmf{dbl_plain}{v1,o2}\fmfv{decor.shape=cross,decor.size=3mm}{o2}
\fmfdot{v2}
\fmflabel{$j$}{v2}\fmf{dbl_plain}{o2,v2}
\end{fmfgraph*}.

\item The integration of the product $(\tilde{D}\bo u)_i(\tilde{D}\bo u)_j$ 
over $\bo u$
corresponds to the contraction of two crossed double straight lines 
coming out of the vertices $i$,$j$.

This produces a matrix element $[B(B+bI)^{-1}B^T]_{i j}$ \,\,\,\,\,\,\ 
\begin{fmfgraph*}(20,1)\fmfpen{thin}
\fmfleft{v1}\fmfright{v2}\fmfdot{v1}
\fmflabel{$i$}{v1}\fmf{dbl_plain}{v1,o2}\fmfv{decor.shape=cross,decor.size=3mm}{o2}
\fmf{dbl_plain}{o2,o3}\fmfv{decor.shape=cross,decor.size=3mm}{o3}
\fmfdot{v2}
\fmflabel{$j$}{v2}\fmf{dbl_plain}{o3,v2}
\end{fmfgraph*}.

\item the integration of the product $\tilde{\gamma}_i\tilde{\gamma}_j$ 
over $\tilde{\bo \gamma}$ corresponds to the contraction of two double 
wiggly lines coming out of vertices $i$,$j$.

This produces a matrix element 
$\tilde{D}_{i j}$ \,\,\,\,\,\,\,\
\begin{fmfgraph*}(20,1)\fmfpen{thin}\fmfleft{v1}\fmfright{v2}\fmfdot{v1}
\fmflabel{$i$}{v1}
\fmfdot{v2}\fmflabel{$j$}{v2}\fmf{dbl_wiggly}{v1,v2}\end{fmfgraph*}.

\item The integration of the product $(J\bo x)_i(J\bo x)_j$ 
over $\bo x$
corresponds to the contraction of two dotted lines 
coming out of the vertices $i$,$j$.

This produces a matrix element $[JCJ^T]_{i j}$ \,\,\,\,\,\,\ 
\begin{fmfgraph*}(20,1)\fmfpen{thin}
\fmfleft{v1}\fmfright{v2}\fmfdot{v1}
\fmflabel{$i$}{v1}\fmf{dots}{v1,v2}
\fmfdot{v2}
\fmflabel{$j$}{v2}
\end{fmfgraph*}.
\end{enumerate}

Eq.(\ref{final_equi2}) can be written in this formalism:

\begin{eqnarray}
&&\left\langle H(\bo V|\bo x)\right\rangle_x=\left\langle H_0
(\bo V|\bo x)\right\rangle_x
+\frac{1}{2}\left\langle\left\langle\left\langle\left\langle
\sum_{l=1}^\infty \frac{g_0^l}{l!\,b^l}\sum_{ijk}
\left[\left(\quad\parbox{10mm}{\begin{fmfgraph*}(10,1)\fmfkeep{u}\fmfpen{thin}
\fmfleft{v1}
\fmfright{o1}\fmflabel{$i$}{v1}\fmfdot{v1}\fmf{dbl_plain}{v1,o1}\end{fmfgraph*}}
-\quad\parbox{10mm}{\begin{fmfgraph*}(10,1)\fmfkeep{t_gamma}\fmfpen{thin}
\fmfleft{v1}\fmfright{o1}
\fmflabel{$i$}{v1}\fmfdot{v1}\fmf{dbl_wiggly}{v1,o1}\end{fmfgraph*}}-\quad\parbox{10mm}{
\begin{fmfgraph*}(10,1)\fmfkeep{du}\fmfpen{thin}\fmfleft{v1}\fmfright{o1}
\fmflabel{$i$}{v1}\fmfdot{v1}\fmf{dbl_plain}{v1,o2}\fmfv{decor.shape=cross,decor.size=3mm}
{o2}
\fmf{dbl_plain}{o2,o1}\end{fmfgraph*}}\right)\right.\right.\right.\right.\right.\label{equi_diagr}\\
&&\left.\left.\left.\left.\left.\left(\quad\parbox{10mm}{\fmfreuse{gamma}}
+\quad\parbox{10mm}{\fmfreuse{du}}+\quad\parbox{10mm}{\begin{fmfgraph*}(10,1)
\fmfkeep{jx}\fmfpen{thin}\fmfleft{v1}
\fmfright{o1}\fmflabel{$i$}{v1}\fmfdot{v1}\fmf{dots}{v1,o1}\end{fmfgraph*}}\right)^3
-\frac{g_0}{2}\left(\quad\parbox{10mm}{\fmfreuse{t_gamma}}+\quad\parbox{10mm}{\fmfreuse{dv}}
+\quad\parbox{10mm}{\fmfreuse{jx}}\right)^6\right]^l\right\rangle_{\parbox{5mm}
{\begin{fmfgraph*}(5,1)\fmfkeep{t_gamma_av}
\fmfpen{thin}\fmfleft{v1}\fmfright{o1}
\fmfdot{v1}\fmf{dbl_wiggly}{v1,o1}\end{fmfgraph*}}}\right.\right.\right.\nonumber\\
&&\left.\left.\left. 
\left(\parbox{15mm}{\begin{fmfgraph*}(15,5)
\fmfpen{thin}\fmfleft{i1}\fmfright{o1}
\fmfpoly{full,tension=0.5}{k1,k2,k3,k4}
\fmf{dbl_plain,label=$j$,label.side=right}{i1,k1}
\fmf{dbl_plain,label=$k$,label.side=right}{k3,o1}\end{fmfgraph*}}\right)
\right\rangle_{\parbox{5mm}{\begin{fmfgraph*}(5,1)\fmfkeep{u_av}
\fmfpen{thin}\fmfleft{v1}\fmfright{o1}
\fmfdot{v1}\fmf{dbl_plain}{v1,o1}\end{fmfgraph*}}}\,\,\right\rangle_{\parbox{5mm}
{\begin{fmfgraph*}(5,1)\fmfkeep{du_av}
\fmfpen{thin}\fmfleft{v1}\fmfright{o1}
\fmfdot{v1}\fmf{dbl_plain}{v1,o2}\fmfv{decor.shape=cross,decor.size=3mm}{o2}\fmf{dbl_plain}{o2,o1}
\end{fmfgraph*}}}\,\,\right\rangle_{\parbox{5mm}{\begin{fmfgraph*}(5,1)\fmfkeep{jx_av}
\fmfpen{thin}\fmfleft{v1}\fmfright{o1}
\fmfdot{v1}\fmf{dots}{v1,o1}\end{fmfgraph*}}}+
\left\langle\left\langle\left\langle\sum_{n=2}^{\infty} \frac{(-1)^{n+1}}{n(n-1)}
\left\langle\sum_{l=1}^\infty \frac{g_0^l}{l!\,b^l}\sum_i
\left[\left(\quad\parbox{10mm}{\fmfreuse{u}}-\quad\parbox{10mm}{\fmfreuse{t_gamma}}
-\quad\parbox{10mm}{\fmfreuse{du}}\right)\right.\right.\right.\right.\right.\nonumber\\
&&\left.\left.\left.\left.\left.\left(\quad\parbox{10mm}{\fmfreuse{t_gamma}}+
\quad\parbox{10mm}{\fmfreuse{du}}+\quad\parbox{10mm}{\fmfreuse{jx}}\right)^3
-\frac{g_0}{2}\left(\quad\parbox{10mm}{\fmfreuse{gamma}}+\quad
\parbox{10mm}{\fmfreuse{du}}+\quad\parbox{10mm}{\fmfreuse{jx}}\right)^6\right]^l
\right\rangle_{\parbox{5mm}{\fmfreuse{t_gamma_av}}}^n\,\,
\right\rangle_{\parbox{5mm}{\fmfreuse{u_av}}}\,\,
\right\rangle_{\parbox{5mm}{\fmfreuse{du_av}}}\,\,
\right\rangle_{\parbox{5mm}{\fmfreuse{jx_av}}}.\nonumber
\end{eqnarray}

Eqs.(\ref{entropy_diagr}),(\ref{equi_diagr}) constitute the final expression 
for the mutual information at every order.
Application of the Wick's theorem and of the contraction rules allows to 
analytically derive the contribution to the mutual information at each order 
in $\bo g$.

The first order approximation in $g_0$, studied 
in \cite{vale+01a}, can be easily obtained from eqs.(\ref{equi_diagr}),(\ref{entropy_diagr}), 
putting $l=1$ and neglecting the double summation over the 
indices $l$ 
and $n$, which gives corrections only at orders higher than first one in $g_0$.

\end{fmffile}

\begin{fmffile}{elka2}

\section{A detailed analysis of the different contributions to the mutual information}

The expansion we have derived allows to investigate how the different 
orders contribute to the mutual information, varying the expansion parameter $g_0$.

The expression of the mutual information at the zeroth order can be easily 
derived from eqs.(\ref{H_0_v}),(\ref{H_0_v_x}):

\be
I_0=H_0(\bo V)-\left\langle H_0(\bo V|\bo x)\right\rangle_x=
\frac{1}{2}\ln det\left(\frac{A+bI}{B+bI}\right);
\label{I_0}
\ee

The first order approximation has been a primary 
object of investigation in \cite{vale+01a}, where a diagrammatic 
interpretation has been provided, as well.
From eqs.(\ref{outfin}),(\ref{eqfin}) it can be shown that the final expression 
of the first order contribution to the 
mutual information can be written as follows:

\begin{equation}
\label{Icub}
I_1=- 3g_0 \sum_{i j} A_{i i}[[A+bI]^{-1}_{i j}A_{i j} - [B+bI]^{-1}_{i j}B_{ij}] ,
\label{I_1}
\end{equation}
Further details about the derivation of the first order approximation can be found in \cite{vale+01a}.

We now focus on the second order contributions.
Let us reconsider eqs.(\ref{entropy_diagr}) and (\ref{equi_diagr}). It is 
clear that the expansion in $l$, $n$ is not a direct expansion in $g_0$:
the first order in $l$ contains both first and second order terms in $g_0$.
Therefore one must be careful and extract all the second order terms in $g_0$ 
from the proper powers in $l,n$.
In particular, in the expansion for both, the output entropy and the equivocation, 
one has to put $l=1,2$ in the first sum and $n=2,l=1$ in the second sum, and then 
retain only the second order terms.
It can be shown that:

\begin{eqnarray}
&&H_2(\bo V)=\frac{g_0^2}{b^2}\left\{\frac{1}{4}\left\langle\left\langle\left\langle
\sum_{ijk}
\left(\quad\parbox{10mm}{\begin{fmfgraph*}(10,1)\fmfkeep{V}\fmfpen{thin}
\fmfleft{v1}
\fmfright{o1}\fmflabel{$i$}{v1}\fmfdot{v1}\fmf{plain}{v1,o1}\end{fmfgraph*}}
-\quad\parbox{10mm}{\begin{fmfgraph*}(10,1)\fmfkeep{gamma}\fmfpen{thin}
\fmfleft{v1}\fmfright{o1}
\fmflabel{$i$}{v1}\fmfdot{v1}\fmf{photon}{v1,o1}\end{fmfgraph*}}-\quad\parbox{10mm}{
\begin{fmfgraph*}(10,1)\fmfkeep{dv}\fmfpen{thin}\fmfleft{v1}\fmfright{o1}
\fmflabel{$i$}{v1}\fmfdot{v1}\fmf{plain}{v1,o2}\fmfv{decor.shape=cross,decor.size=3mm}{o2}
\fmf{plain}{o2,o1}\end{fmfgraph*}}\right)^2\left(\quad\parbox{10mm}{\fmfreuse{gamma}}
+\quad\parbox{10mm}{\fmfreuse{dv}}\right)^6\right.\right.\right.\right.\nonumber\\
&&\left.\left.\left.\left.-b \left(\quad\parbox{10mm}{\fmfreuse{gamma}}+\quad\parbox{10mm}{\fmfreuse{dv}}\right)^6
\right\rangle_{\parbox{5mm}{\begin{fmfgraph*}(5,1)\fmfkeep{gamma_av}
\fmfpen{thin}\fmfleft{v1}\fmfright{o1}
\fmfdot{v1}\fmf{photon}{v1,o1}\end{fmfgraph*}}} 
\!\!\!\!
\left(\parbox{15mm}{\begin{fmfgraph*}(15,5)
\fmfpen{thin}\fmfleft{i1}\fmfright{o1}
\fmfpoly{shade,tension=0.5}{k1,k2,k3,k4}
\fmf{plain,label=$j$,label.side=right}{i1,k1}
\fmf{plain,label=$k$,label.side=right}{k3,o1}\end{fmfgraph*}}\right)
\right\rangle_{\parbox{5mm}{\begin{fmfgraph*}(5,1)\fmfkeep{V_av}
\fmfpen{thin}\fmfleft{v1}\fmfright{o1}
\fmfdot{v1}\fmf{plain}{v1,o1}\end{fmfgraph*}}}\,\right\rangle_{\parbox{5mm}
{\begin{fmfgraph*}(5,1)\fmfkeep{dv_av}
\fmfpen{thin}\fmfleft{v1}\fmfright{o1}
\fmfdot{v1}\fmf{plain}{v1,o2}\fmfv{decor.shape=cross,decor.size=3mm}{o2}\fmf{plain}{o2,o1}
\end{fmfgraph*}}}\right.\nonumber\\
&&\left.-\frac{1}{2}
\left\langle\left\langle
\left\langle \sum_i
\left(\quad\parbox{10mm}{\fmfreuse{V}}-\quad\parbox{10mm}{\fmfreuse{gamma}}
-\quad\parbox{10mm}{\fmfreuse{dv}}\right)\left(\quad\parbox{10mm}{\fmfreuse{gamma}}+
\quad\parbox{10mm}{\fmfreuse{dv}}\right)^3
\right\rangle_{\parbox{5mm}{\fmfreuse{gamma_av}}}^2\,\,
\right\rangle_{\parbox{5mm}{\fmfreuse{V_av}}}\,\,
\right\rangle_{\parbox{5mm}{\fmfreuse{dv_av}}}\right\}
\end{eqnarray}

\begin{eqnarray}
&&\left\langle H_2(\bo V|\bo x)\right\rangle_x=\nonumber\\
&&\frac{g_0^2}{b^2}\left\{
\frac{1}{4}\left\langle\left\langle\left\langle\left\langle\sum_{ijk}
\left(\quad\parbox{10mm}{\begin{fmfgraph*}(10,1)\fmfkeep{u}\fmfpen{thin}
\fmfleft{v1}
\fmfright{o1}\fmflabel{$i$}{v1}\fmfdot{v1}\fmf{dbl_plain}{v1,o1}\end{fmfgraph*}}
-\quad\parbox{10mm}{\begin{fmfgraph*}(10,1)\fmfkeep{t_gamma}\fmfpen{thin}
\fmfleft{v1}\fmfright{o1}
\fmflabel{$i$}{v1}\fmfdot{v1}\fmf{dbl_wiggly}{v1,o1}\end{fmfgraph*}}-\quad\parbox{10mm}{
\begin{fmfgraph*}(10,1)\fmfkeep{du}\fmfpen{thin}\fmfleft{v1}\fmfright{o1}
\fmflabel{$i$}{v1}\fmfdot{v1}\fmf{dbl_plain}{v1,o2}\fmfv{decor.shape=cross,decor.size=3mm}
{o2}\fmf{dbl_plain}{o2,o1}\end{fmfgraph*}}\right)^2\left(\quad\parbox{10mm}{\fmfreuse{gamma}}
+\quad\parbox{10mm}{\fmfreuse{du}}+\quad\parbox{10mm}{\begin{fmfgraph*}(10,1)
\fmfkeep{jx}\fmfpen{thin}\fmfleft{v1}
\fmfright{o1}\fmflabel{$i$}{v1}\fmfdot{v1}\fmf{dots}{v1,o1}\end{fmfgraph*}}\right)^6 
\right.\right.\right.\right.\right.\nonumber\\
&&\left.\left.\left.\left.\left.
-b\left(\quad\parbox{10mm}{\fmfreuse{t_gamma}}+\quad\parbox{10mm}{\fmfreuse{du}}+
\quad\parbox{10mm}{\fmfreuse{jx}}\right)^6
\right\rangle_{\parbox{5mm}
{\begin{fmfgraph*}(5,1)\fmfkeep{t_gamma_av}
\fmfpen{thin}\fmfleft{v1}\fmfright{o1}
\fmfdot{v1}\fmf{dbl_wiggly}{v1,o1}\end{fmfgraph*}}}
\left(\parbox{15mm}{\begin{fmfgraph*}(15,5)
\fmfpen{thin}\fmfleft{i1}\fmfright{o1}
\fmfpoly{full,tension=0.5}{k1,k2,k3,k4}
\fmf{dbl_plain,label=$j$,label.side=right}{i1,k1}
\fmf{dbl_plain,label=$k$,label.side=right}{k3,o1}\end{fmfgraph*}}\right)
\right\rangle_{\parbox{5mm}{\begin{fmfgraph*}(5,1)\fmfkeep{u_av}
\fmfpen{thin}\fmfleft{v1}\fmfright{o1}
\fmfdot{v1}\fmf{dbl_plain}{v1,o1}\end{fmfgraph*}}}\,\,\right\rangle_{\parbox{5mm}
{\begin{fmfgraph*}(5,1)\fmfkeep{du_av}
\fmfpen{thin}\fmfleft{v1}\fmfright{o1}
\fmfdot{v1}\fmf{dbl_plain}{v1,o2}\fmfv{decor.shape=cross,decor.size=3mm}{o2}\fmf{dbl_plain}{o2,o1}
\end{fmfgraph*}}}\,\,\right\rangle_{\parbox{5mm}{\begin{fmfgraph*}(5,1)\fmfkeep{jx_av}
\fmfpen{thin}\fmfleft{v1}\fmfright{o1}
\fmfdot{v1}\fmf{dots}{v1,o1}\end{fmfgraph*}}}\right.\nonumber\\
&&\left.-\frac{1}{2}\left\langle\left\langle\left\langle
\left\langle\sum_i
\left(\quad\parbox{10mm}{\fmfreuse{u}}-\quad\parbox{10mm}{\fmfreuse{t_gamma}}
-\quad\parbox{10mm}{\fmfreuse{du}}\right)\right.\right.\right.\right.\right.\nonumber\\
&&\left.\left.\left.\left.\left.\left(\quad\parbox{10mm}{\fmfreuse{t_gamma}}+
\quad\parbox{10mm}{\fmfreuse{du}}+\quad\parbox{10mm}{\fmfreuse{jx}}\right)^3
\right\rangle_{\parbox{5mm}{\fmfreuse{t_gamma_av}}}\,\,
\right\rangle_{\parbox{5mm}{\fmfreuse{u_av}}}\,\,
\right\rangle_{\parbox{5mm}{\fmfreuse{du_av}}}\,\,
\right\rangle_{\parbox{5mm}{\fmfreuse{jx_av}}}\right\},
\end{eqnarray}
where we have extracted the terms which are order $g_0^2$.

The computation of all the second order diagrams is very long, since it involves sixth 
powers of terms and $4$ different types of contractions, both in the equivocation and 
in the output entropy. 
Yet the calculation results much simplified in some limits.  

In particular, let us consider the case where $g_0\ll b$ but $b$ is not too small.
Both the equivocation and the output entropy contain terms of order $g_0^2/b^2$ up to 
$g_0^2b^2$. In fact, looking at the contraction rules given above, one can notice that 
the contraction of two wiggly lines produces a matrix element $D_{ij}$, where the matrix 
$D$ is order $b$, 
while the other contractions produce elements which are order 1 with respect to $b$.
Therefore the dominating contributions in the limit when $b$ is not 
too small are given by the diagrams where $3$ or $4$ wiggly loops appear. 
Keeping only these diagrams, after having performed all the contractions, 
it can be shown that the dominating second order contribution to the output entropy 
is given by the following sum of diagrams: 

\be
H_2(\bo V)\simeq -\frac{9}{2}\frac{g_0^2}{b^2}\left[2\sum_{ij}\quad
\parbox{25mm}{\begin{fmfgraph*}(15,25)\fmfpen{thin}\fmfstraight
\fmfleft{i1,i2,i3}\fmfright{o1,o2,o3}\fmf{phantom}{i1,v1,o1}\fmf{phantom}{i3,v3,o3}
\fmf{phantom}{i2,v2,o2}
\fmffreeze\fmf{wiggly,right=0.5,tension=0.3}{i1,i2,i1}
\fmf{wiggly,left=0.5,tension=0.3}{i3,i2,i3}\fmflabel{$i$}{i2}\fmflabel{$j$}{o2}
\fmf{wiggly,right=0.5,tension=0.3}{o1,o2,o1}
\fmf{wiggly,left=0.5,tension=0.3}{o3,o2,o3}\fmfdot{i2}\fmfdot{o2}
\end{fmfgraph*}}\right],
\ee
which corresponds to the following analytical expression:
\be
H_2(\bo V)\simeq -\frac{9}{2}g_0^2b^2
\left[Tr\left(A\left(A+bI\right)^{-1}\right)^2\right]^2.
\ee
Under the same assumption the second order contribution to the equivocation can also 
be expressed as 
a simple sum of diagrams:
\be
\left\langle H(\bo V|\bo x)\right\rangle_x\simeq 
-\frac{g_0^2}{b^2}\sum_i\left[\frac{15b}{4}\quad
\parbox{25mm}{\begin{fmfgraph*}(25,25)\fmfpen{thin}
\fmfleft{i1,i2,i3}\fmfright{o1,o2,o3}\fmf{phantom}{i3,v3,o3}\fmf{phantom}{i2,v2,o2}
\fmffreeze
\fmf{dbl_wiggly,right=0.5,tension=0.3}{i2,v2,i2}
\fmfdot{v2}\fmfv{label=$i$,label.angle=90}{v2}
\fmf{dbl_wiggly,left=0.5,tension=0.3}{v2,o2,v2}\fmf{phantom}{v1,v2,v1}
\fmf{dbl_wiggly,right=0.5,tension=0.3}{v3,v2,v3}
\end{fmfgraph*}}
+7\quad\parbox{25mm}{\begin{fmfgraph*}(25,25)\fmfkeep{4petals}\fmfpen{thin}
\fmfleft{i1,i2,i3}\fmfright{o1,o2,o3}\fmf{phantom}{i1,v1,o1}\fmf{phantom}{i3,v3,o3}
\fmffreeze\fmf{dbl_wiggly,right=0.5,tension=0.3}{i2,v2,i2}
\fmfdot{v2}\fmfv{label=$i$,label.angle=90}{v2}
\fmf{dbl_wiggly,left=0.5,tension=0.3}{v2,o2,v2}
\fmf{dbl_wiggly,right=0.5,tension=0.3}{v3,v2,v3}
\fmf{dbl_wiggly,right=0.5,tension=0.3}{v1,v2,v1}
\end{fmfgraph*}}
\right].
\ee
After some elementary manipulation of the matrices it is easy to show that 
the analytical expression for the equivocation reads:
\be
\left\langle H(\bo V|\bo x)\right\rangle_x\simeq
-g_0^2b^2\left[\frac{15}{4}Tr \left(B\left(B+bI\right)^{-1}\right)^3+
7\,Tr \left(B\left(B+bI\right)^{-1}\right)^4\right];
\ee
As a whole, the second order contribution to the mutual information in 
the limit when $b$ is not too small can be written:
\be
I_2\simeq 3g_0^2b^2\left(\frac{5}{4}Tr \left(B\left(B+bI\right)^{-1}\right)^3+
\frac{7}{3}Tr \left(B\left(B+bI\right)^{-1}\right)^4-
\frac{3}{2}\left[Tr\left(A\left(A+bI\right)^{-1}\right)^2\right]^2\right);
\label{I_2}
\ee

Since the evaluation of the mutual information has been carried out for a generic 
connectivity matrix $\{J_{ij}\}$ it is obvious that both the total information 
value and its order-specific contributions will depend on the structure of the 
connectivities. 
Even without aiming at a generalization we can in any case provide a 
quantification of the different contributions 
to the mutual information restricting ourselves to the specific case where the connectivities 
linking each output neuron to the input ones are drawn from a gaussian distribution 
with mean zero and standard deviation $1/\sqrt{N}$. 
As it is known from the theory of spin glasses and neural networks, this renormalization 
ensures that the local field $h_i$ acting on each unit $i$ is finite when $N$ is large.
 
\begin{figure}
\centerline{
\psfig{figure=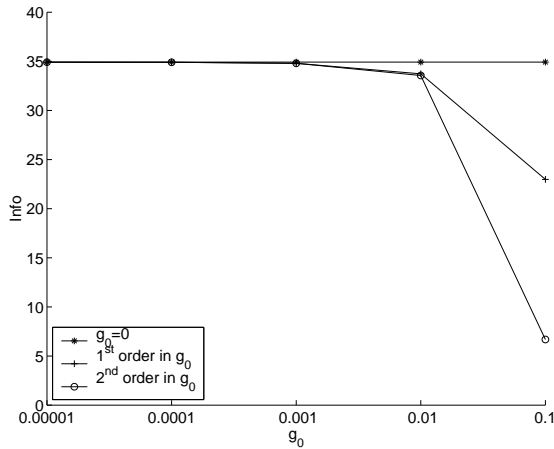,height=6cm,angle=0}}
\caption{Mutual information as a function of the parameter $g_0$, comparing the zeroth, 
first and second order approximation. $N=100$;$b_0=0.1$; $b=0.6$. The correlation matrix $C$ is 
equal to the identity and the connectivities $J_{ij}$ linking each output unit 
$i$ to the input neurons $j$ are randomly chosen from a Gaussian 
distribution with mean zero and standard deviation $1/\sqrt{N}$} 
\label{elka}  
\end{figure}

Fig.(\ref{elka}) shows the mutual information in zeroth, first and second order 
approximation for increasing values of $g_0$. Two main observations stem from the 
analysis of the curves:
\begin{itemize}
\item both the first and the second order contributions in the non-linearity lower the 
information value with respect to the linear network.
\item our approximations start to lose their validity for values 
of $g_0$ larger than $0.01$; this does not automatically mean that one should add higher 
order contributions, since we have neglected second order contributions with powers of $b$ 
lower than $b^2$: when $g_0$ becomes not much smaller than $b$ one should probably include 
the other contributions, at second order in $g_0$.
\end{itemize} 

\begin{figure}[h]
\centerline{
\psfig{figure=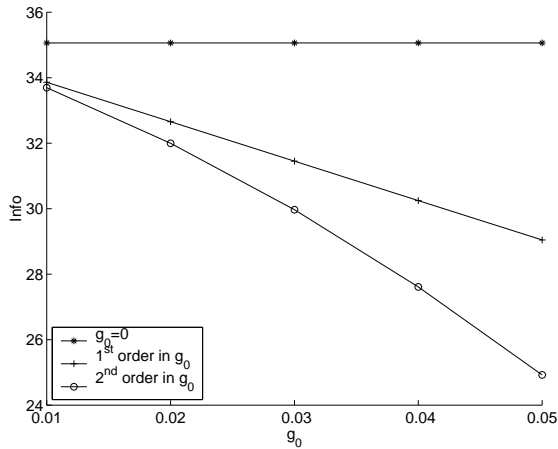,height=6cm,angle=0}}
\caption{Mutual information as a function of the parameter $g_0$, as in fig.(\ref{elka}).
$N=100$;$b_0=0.1$; $b=0.6$. The correlation matrix $C$ is 
equal to the identity and the connectivities $J_{ij}$ linking each output unit 
$i$ to the input neurons $j$ are randomly chosen from a Gaussian 
distribution with mean zero and standard deviation $1/\sqrt{N}$} 
\label{elka_det}  
\end{figure}

Fig.(\ref{elka_det}) shows a detail of the previous plot, where the linear and quadratic 
fit in $g_0$ are more evident.

\begin{figure}[h]
\centerline{
\psfig{figure=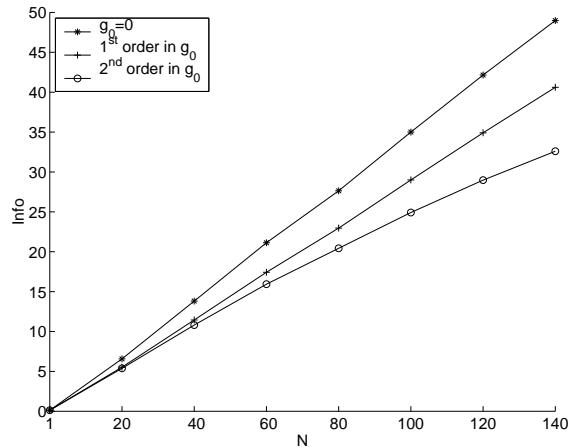,height=6cm,angle=0}}
\caption{Mutual information as a function of the population size $N$, comparing the zeroth, 
first and second order approximations in $g_0$. 
$g_0=0.05$;$b_0=0.1$; $b=0.6$. The correlation matrix $C$ is 
equal to the identity and the connectivities $J_{ij}$ linking each output unit 
$i$ to the input neurons $j$ are randomly chosen from a Gaussian 
distribution with mean zero and standard deviation $1/\sqrt{N}$} 
\label{elka_N}  
\end{figure}

Fig.(\ref{elka_N}) shows the mutual information as a function of the population size $N$.
As it is clear from eqs.(\ref{I_0}),(\ref{I_1}),(\ref{I_2}), the zeroth order approximation 
is linear in $N$, and so is also the first order contribution, since it is the difference 
between two scalar products of $N$-dimensional vectors.
On the other hand the second order contribution is roughly quadratic in $N$: in fact a numerical 
check of the three different contributions shows that the first two terms, which are linear in $N$ 
(traces of $N$-dim matrices) are three orders of magnitudes smaller than the third term, which 
is quadratic in $N$ (square of a trace of an $N$-dim matrix).

\section{Conclusions}
We have presented here a systematic approach to quantify the effect of small 
non-linearities in the transfer function on the information transferred 
by a two layer network of analogue units.
We have derived a perturbative expansion in the non-linearity parameter $g_0$, 
providing an elegant interpretation in terms of Feynman diagrams.

In a previous report \cite{vale+01a} we had already quantified the contribution to the information 
at first order in $g_0$. 
Here we have extended the previous results providing an analytical expression to calculate 
the contributions at each order in perturbation theory. Moreover 
our method can be easily applied to any structure of the connectivities, with no 
restriction to a specific architecture.

As an example, we have quantified the zeroth, first and second order contributions 
to the information in the case of random, Gaussian distributed couplings and in the 
limit when the output noise $b$ is not small.
We have found that the main effect of the non-linearity for this particular architecture 
is a loss in information, detected both at first and at second order in $g_0$.
This result is in agreement with previous investigations \cite{vale+01c}, 
where it has been 
shown that two main causes of information loss in a two-layer network with random gaussian couplings 
are a non-linearity in the transfer function, like the presence of a threshold, and a large 
output noise. 

The detailed analysis we have presented here applies to the particular 
case of cubic 
non-linearities. Yet, as already remarked in \cite{vale+01a}, our 
method can be 
easily adapted to any non-linearity of a generic power $2k+1$: it is enough 
to replace the third and sixth powers appearing in eqs.(\ref{entropy_diagr}) and 
(\ref{equi_diagr}) respectively with powers $2k+1$
and $4k+2$.

Finally this 
allows to deal with highly non-linear functions, like a hyperbolic tangent, which has often 
been proposed in modelling realistic neural systems \cite{Ami+89}.


\section*{Acknowledgments}
E.K. highly acknowledges the financial help from the Grant
BFM2001-291-C02-01 from the Spanish Ministry of Science and Technology.
V.D.P. thanks D.Boll\'e for his support and suggestions. This work has been 
partially supported by Fund of Scientific Research, Flanders-Belgium.

\end{fmffile}

\end{document}